\newif\ifchapter
\chaptertrue

\ifchapter
\documentclass[graybox]{svmult}
\else
\documentclass[useAMS,usenatbib]{mn2e}
\fi

\newcommand{\comment}[1]{}

\ifchapter
\usepackage{mathptmx}       
\usepackage{helvet}         
\usepackage{courier}        
\usepackage{type1cm}        
\usepackage{makeidx}         
\usepackage{graphicx}        
\usepackage{multicol}        
\usepackage[bottom]{footmisc}

\makeindex             
\else
\usepackage{times}
\usepackage[dvips]{graphicx}
\usepackage{color}
\usepackage[T1]{fontenc}
\usepackage{ae,aecompl}
\fi

\usepackage{aas_macros}
\usepackage{amssymb}
\usepackage{natbib} 
\usepackage{abrev} 

\begin{document}

\ifchapter
\title*{Bulge Formation via Mergers in Cosmological Simulations}
\author{Alyson Brooks and Charlotte Christensen}
\institute{Alyson Brooks \at Rutgers University, 136 Frelinghuysen Rd, Piscataway, NJ 08854 United States, \email{abrooks@physics.rutgers.edu}
\and Charlotte Christensen \at Grinnell College, 1116 Eighth Ave, Grinnell, IA 50112 United States, \email{christenc@grinnell.edu}}
%
%
\fi

\ifchapter
\maketitle
\abstract*{Abstract (10--15 lines long)}
\abstract{
The latest generation of cosmological simulations are on the verge of being able to resolve the structure of bulges for the first time.  
Hence, we review the current state of bulge formation in cosmological simulations, and discuss open questions that can be addressed in the near future by simulators, with a particular focus on merger-driven bulge growth.
Galaxy mergers have long been assumed to produce classical bulges in disk galaxies.  Under this bulge-formation model, though, the high rates of mergers in Cold Dark Matter galaxy formation theory predict many more classical bulges than are observed.  Furthermore, simulations of galaxy formation continue to generally produce too massive of bulges.
Feedback offers a promising avenue for reducing merger-driven bulge growth by maintaining high gas fractions in galaxies and ejecting low-angular momentum gas driven to the centers of galaxies.
After reviewing the results of relevant research that has been published to date, we use cosmological simulations to explore the ability of feedback to reduce or even prevent bulge growth during mergers.
In dwarf galaxies, mergers actually reduce the central concentration of galaxies as the induced burst of star formation drives out low angular momentum material.
This result shows the potential for feedback to reduce central mass growth. 
However, we also demonstrate that it is very difficult for current stellar feedback models to reproduce the small bulges observed in more massive disk galaxies like the Milky Way.
We argue that feedback models need to be improved, or an additional source of feedback such as AGN is necessary to generate the required outflows.
}
\fi

\section{Introduction}
Galaxy simulators have made significant progress in recent years in being able to simulate realistic disk galaxies that match a range of observed properties \citep[e.g.,][]{Brook2012d, Aumer2013}.  This success has included progress toward resolving the central regions of galaxies and forming more realistic bulges \citep{Christensen14}. 
However, galaxy simulations are, if anything, {\em too} successful at forming bulges.
Even in state-of-the-art simulations, galaxy bulges remain on the massive end of the observed range and the implication is that they also form to easily.

Galaxy simulators spent their formative years trying to understand and overcome the sources of overcooling \citep[e.g.,][]{sn99,ns00, Abadi2003c, governato04, cecilia2}, which leads to an overly dense concentration of mass in the central regions of simulated galaxies compared to observations.  Cosmological simulations achieved a major step forward within the past five years when they were able to create bulgeless dwarf disk galaxies for the first time \citep{Governato2010, Teyssier2013}.  However, the same star formation and feedback scheme that leads to bulgeless dwarf disks still tends to lead to overly massive stellar bulges in halos of $\sim 10^{11}$ M$_{\odot}$ and higher \citep[e.g.][]{Christensen14}.  
A simple interpretation of this trend suggests that more feedback may be required at higher galaxy masses.  Unfortunately, while the most recent simulations have shown that more feedback can indeed lead to smaller bulges in Milky Way-mass galaxies \citep{Aumer2013, Mollitor2014}, the additional feedback creates new challenges compared to observations \citep{Roskar2013, Aumer2014}.

Fully cosmological simulations are the best tool for capturing the detailed merger history of galaxies, and thus the best tool to study the impact of mergers on bulge growth.
However, analyzing the growth of bulges in cosmological simulations is difficult simply because, until recently, the structure of the bulges was unresolved in these simulations.
The highest resolution simulations of Milky Way-mass galaxies \citep[e.g.][]{Hopkins2013d} are only now beginning to have resolved bulges, but over the next few years this sample will greatly expand.
Given that advances in computing will soon allow for higher resolution studies of stellar bulges formed in a cosmological context, we outline in this review the current challenges that will need to be addressed.  In particular, we focus on the role of mergers in forming the bulges of galaxies that are disk-dominated at $z=0$.

We begin in Section 2 by summarizing the observations that inform our investigations, and in Section 3 discuss the current formation ideas that explain bulge properties.  In Section 4 we highlight the current theoretical challenges that emerge when favored bulge formation mechanisms are imposed in a Cold Dark Matter galaxy formation context.  In Section 5, we review the state of cosmological bulge formation, including the limited number of studies on the origin of bulges have so far been carried out with cosmological simulations. We discuss the challenges that remain to forming realistic bulges.  In Section 6 we use simulations to emphasize bulge trends with galaxy mass, and to point out where simulated trends break down in comparison to observations.  We conclude in Section 7 with a discussion of possible solutions to form realistic bulges in mergers. 

\section{Observational Properties of Bulges}
\label{sec:0}
Bulges are frequently divided into two main types: classical bulges and pseudobulges.  
Classical bulges generally are more spherically symmetric, are supported by velocity dispersion, have isotropic velocity distributions, and have older stellar populations, while pseudobulges are distinguished by their more disk-like features \citep{Kormendy1993, Andredakis1994}.

Classical bulges resemble elliptical galaxies in a number of ways.  AS \citet{Renzini1999} stated: ``It appears legitimate to look at bulges as ellipticals that happen to have a prominent disk around them."  \citet{KK04} noted that this statement seems to apply only to classical bulges.  One of the most obvious ways that classical bulges resemble elliptical galaxies is that the surface brightness profiles of classical bulges can be fit with a S\'{e}rsic profile where $n$ is generally equal to or larger than 2:
\begin{equation}
I(r) = I_0 \, exp[(-{\frac{r}{r_0}})^{(1/n)} ]
\label{eq:sb}
\end{equation}
where $I_0$ is the central intensity and $r_0$ is the scaling radius.
The stellar populations in both classical bulges and ellipticals tend to be older \citep{Moorthy2006}, with the stars being typically $\alpha$-enhanced, indicating that the formation happened rapidly \citep[on a timescale short enough that SN\,Ia did not yet contribute iron,][]{Weiss1995, Matteucci2006, Ganda2007, Peletier2007}.  Finally, there is an argument that many of the scaling relations that elliptical galaxies are observed to follow seem to also be followed by classical bulges \citep{Fisher2008,Kormendy2012}, including the Faber-Jackson relation \citep[a luminosity -- velocity dispersion relation,][]{Faber1976}, the Kormendy relation \citep[a size -- luminosity relation,][]{Kormendy1977}, and the fundamental plane, which relates size, velocity, and luminosity. It should be noted, however, that classical bulges do not appear to follow all elliptical galaxy scaling relations: \citet{Gadotti2009} and \citet{Laurikainen2010} found that the bulge mass-size relation for classical bulges was offset from that for elliptical galaxies.

The bulges of many disk galaxies, however, are not spherically symmetric, are rotationally dominated, and have shallower surface brightness profiles ($n \lesssim 2$).
These are known as pseudobulges.
Pseudobulges can be further divided into {\em disky pseudobulges}, which generally have on-going star formation \citep{Fisher2006} and nuclear bars, spirals or rings \citep[e.g.][]{Fisher2006}, and {\em Boxy/Peanut (B/P)} bulges, which are made up of older stellar populations.  
These B/P bulges have  characteristically boxy or peanut isophotal shapes when viewed edge-on and barlens isophotal shapes when viewed face-on and are generally associated with bars \citep{Athanassoula2005}. Boxy bulges show cylindrical rotation (i.e., their rotational velocity is constant with height above the mid plane), unlike classical bulges. 

A review of the observational properties of pseudobulges can be found in \citet{KK04}. 
To summarize, though:
\begin{itemize}
\item Classical bulges have rounder and more spherically symmetric morphology while pseudobulges may have either disky or boxy morphology.
\item The stellar kinematics of classical bulges are generally more dominated by dispersion while in pseudobulges they are more dominated by rotation. 
\item Most classical bulges have S\'{e}rsic indices $n \geq 2$ while most pseudobulges have $n < 2$.
\item Classical bulges tend to follow the Faber-Jackson relation and lie along the fundamental plane. Pseudobulges are frequently low-$\sigma$ outliers from the Faber-Jackson relation.  They may also have fainter effective surface brightness at their effective radii than would be expected from the fundamental plane.
\item Galaxies with a bulge-to-total ($B/T$) ratio $\geq 0.5$ almost always contain a classical bulge.
\item Pseudobulges may contain embedded bars, nuclear rings or, in the case of disky pseudobulges, spiral structure.
\item Disky pseudobulges may also contain young stars, gas, and dust, even when the galaxy is not undergoing a merger.
\end{itemize}
Note that there can be significant overlap in the properties of classical bulges and pseudobulges so the classification between the two should be done by looking at multiple indicators, when possible.

 \subsection{Population Studies of Classical and pseudobulges}
It has become increasingly clear that pseudobulges are ubiquitous throughout the Universe.
\citet{Fisher2011} determined that within a 11 Mpc sphere, 80\% of galaxies with a stellar mass of $10^9 \Msun$ or greater are either bulgeless or contain a pseudobulge (which they defined as any bulge with a low S\'ersic index, i.e., including all types of pseudobulges).
Similar conclusions were reached by \citet{Kormendy2010} within a 8 Mpc sphere.
Current best estimates for the frequency of specifically B/P-shaped pseudobulges are between 20\% \citep{Yoshino2014} and 40\% \citep{Lutticke2000} of spiral galaxies.
Classical bulges are more common in more massive galaxies.
For instance, they are found in the majority of galaxies with stellar mass greater than $10^{10.5} \Msun$ and earlier type galaxies show more classical bulges than late-type galaxies \citep{Andredakis1995, KK04}. 
Similarly, \citet{Fisher2008} found that bluer galaxies were more likely to host disky pseudobulges, while red galaxies were more likely to host classical bulges. In contrast to disky pseudobulges, B/P bulges are slightly more common in earlier-type spiral galaxies that have bars \citep{Laurikainen2014a,Yoshino2014}.

Despite classical bulges being most common in more massive galaxies, even giant ($V_{circ} \geq 200$ km/s) Sc-Scd galaxies frequently lack a classical bulge \citep{Kormendy2010}.
Perhaps most notable of this group is the Milky Way itself.
Photometric \citep{Blitz1991,Dwek1995} and kinematic \citep{Howard2009, Shen2010} evidence shows that the Milky Way's bulge has the boxy shape characteristic of bars seen edge-on \citep{Combes1990,Raha1991,Athanassoula2005}.  
While it is possible that an additional classical bulge component could be hidden by the rest of the bulge, kinematic and metallicity data limit any classical bulge contribution to \textless10\% the disk mass \citep{Shen2010, DiMatteo2014}.

Not only are classical bulges more common in higher mass galaxies, they are also less common in low-density environments than in high-density ones \citep{Kormendy2010}.
Furthermore, the B/T ratios in low-density environments also tend to be smaller \citep{Kautsch2009}.
Theorists must, therefore, explain not only the abundance of massive pure disk galaxies but also their environmental dependency.

\section{Theoretical models for the formation of bulges}
\label{sec:1}
A common view is that classical bulges are formed primarily in mergers of galaxies, while pseudobulges are formed by processes internal to the galaxy\footnote{One possible internal process for bulge formation we will not discuss here is the build-up of bulges from clumpy disks.  In this model,  large clumps in massive disks at high $z$ may migrate to the center to form bulges \citep{Genzel2008, Bournaud2013, Dekel2013, Perez2013}.  We leave a complete discussion of this formation process to chapter 6.2 (Bournaud, 2015). However, it should be noted that the outcomes of gas-rich clumps sinking to the centers of galaxies are not unlike those of gas-rich mergers \citep{Kormendy2013}.}.
Below, we review the commonly accepted paradigms for formation of classical and pseudobulges in turn.

\subsection{Classical Bulges}
The similarities between elliptical galaxies and bulges suggest a common formation origin.
Mergers have been considered likely triggers for elliptical galaxy formation for even longer than they have been considered the source of bulges \citep{Toomre1977}.  
Mergers offer several compatible avenues for bulge growth: the violent relaxation of the primary galaxy stellar component, the accretion and violent relaxation of the secondary galaxy stellar component, and the formation of stars from gas undergoing merger-driven angular momentum loss.
Given the observed similarities between elliptical galaxies and classical bulges, it is not surprising that the merger hypothesis for elliptical galaxies has been extended to the bulges of spiral galaxies.

\subsubsection{Merger formed Ellipticals?}
Do elliptical galaxies form in mergers? 
Simulations of disk galaxy mergers show that the disruption and violent relaxation \citep{LyndenBell1967} of the stars results in the formation of a spheroid with the density distribution similar to a de Vaucouleurs ($r^{1/4}$) profile \citep{Barnes1988, Hernquist1992}. \citet{Hopkins2008a} used dissipational mergers of disk galaxies to show that the stars that existed prior to the merger were redistributed into a spheroidal component that could be fit by a S\'{e}rsic profile with $n > 2.5$, in agreement with the observed surface brightness profiles of ellipticals \citep{Kormendy2009}.  
Additionally, many elliptical galaxies are observed to have excess light at their centers compared to what would be extrapolated from a S\'ersic fit, in what is termed a "cuspy'' profile \citep{Kormendy2009,Krajnovic2013}.
Such cuspy profiles are thought to be the signature of a central starburst resulting from a dissipational merger \citep{Mihos1994, Kormendy1999}.
This observation is consistent with the trend for cuspy ellipticals to be less massive \citep{Kormendy2009} and faster rotators \citep{Krajnovic2013} than ellipticals without excess light.
On the other hand, dissipational mergers may not be necessary to produce these observations: simulations of direct collisionless collapse have also been shown to result in a density profile that follows roughly $r^{1/4}$, as well as decreasing line-of-sight velocity dispersions and increasing velocity anisotropy \citep{Hozumi2000, Trenti2005}, which are all trends observed in ellipticals.  

Certainly the old stellar populations in ellipticals require that they form at a time when mergers were more frequent than today.  Despite this, there has been some questions as to whether mergers alone are enough to reproduce the Fundamental Plane, Faber--Jackson, and Kormendy relations observed for elliptical galaxies.  Two merger regimes have been explored: those with gas and those that are purely collisionless.  It has been known for a while that there are problems reproducing phase space and elliptical galaxy scaling relations with collisionless merger alone \citep{Ostriker1980, Carlberg1986}.  For instance, the apparent size growth of ellipticals also cannot be explained by gas-poor mergers in the latter half of the age of the Universe \citep{Cimatti2006, Renzini2006, Bundy2007, Scarlata2007a, Nipoti2009}.  However, this leaves open the idea that ellipticals were formed in gas-rich mergers early in the age of the Universe.  Mergers with significant dissipation do a better job of building the scaling relations of ellipticals than gas-poor mergers \citep{Mihos1994, naab07, Jesseit2009} but \citet{Ciotti2007} concluded that some form of initial monolithic collapse was necessary in addition to mergers to match the scaling relations.
More recently, however, both SAMs \citep{Porter2014} and cosmological simulations \citep{Oser2012} have found that a combination of wet and dry mergers and major and minor mergers produced galaxies that followed the observed slope and time evolution of the size-mass relation and, in the former case, the Faber--Jackson relation.
 
Overall, it seems some initial gas-rich collapse and subsequent quenching is required to reproduce the detailed structural properties of ellipticals.  However, it is clear that mergers can redistribute existing stars via violent relaxation into a light profile consistent with observations. Hence, both dissipational collapse and mergers are likely involved in the formation of elliptical galaxies.   Are these also the processes that form classical bulges?

\subsubsection{Merger formed Classical Bulges?}

When gas is neglected, major mergers of disk galaxies tend to result in an elliptical galaxy rather than a disk galaxy, as discussed above.  In the absence of dissipation, one might imagine that the general distribution of stellar orbits in the remnant correspond roughly to their initial radii prior to the merger.  This does not lead to a compact distribution consistent with bulge growth.  On the other hand, minor mergers in the mass ratio range 4:1 to 10:1 can result in disks with bulges \citep{Bournaud2005}.  In that case, though, the bulge growth is dominated by gas flows to the central region of the primary galaxy.  As we will discuss below, \citet{hopkins09} demonstrated that bulge growth is significantly suppressed as gas fraction increases.  \citet{Hopkins2009a} then showed that incorporating this model into the merger histories of halos in $\Lambda$CDM can reproduce the trends in spheroid morphology with galaxy mass.  A major implication of this successful model is that bulge growth is driven by dissipative processes rather than redistribution of stars. 
Indeed, \citep{toomre72} stated:  ``Would not the violent mechanical agitation of a close tidal encounter -- let alone an actual merger -- already tend to bring {\em deep} into a galaxy a fairly {\em sudden} supply of fresh fuel in the form of interstellar material, either from its own outlying disk or by accretion from its partner?''  If channeling of gas in mergers plays a prominent role in bulge formation, then most bulges would appear to be formed in situ to the galaxy.  In fact, this is in agreement with current cosmological simulations of disk galaxies \citep[discussed further in Section~\ref{sec:3},][]{Guedes2013, Okamoto2013, Christensen14, Pillepich2014b}.

Dissipational simulations of mergers result in the build up of a central gas mass \citep[e.g.,][]{Negroponte1983}.
\citet{Hernquist1989} and \citet{Barnes1991} established the theoretical underpinnings for this funneling of gas to the centers of galaxies during mergers.  They showed that gas loses angular momentum during mergers through interactions with induced bars, tidal torques and dynamical friction with the disk stars \citep[see also][]{Noguchi1988, Combes1990a, Barnes1996}. 
While initially the transfer of gas to the centers of galaxies was primarily seen as a way to fuel Active Galactic Nuclei (AGN), \citet{katz92} suggested that bulges form from gas that underwent substantial merger-induced dissipation and from the stars accreted during the merger.
This picture of rapid star formation from suddenly condensed gas is consistent with observations of the blue colors \citep[e.g.][]{Larson1978a}, increased star formation rates \citep{Ellison2008, Jogee2009}, and high central gas densities \citep{Young1984,Sanders1985,Sanders1987} of merging and tidally deformed galaxies. It is also consistent with the apparent funneling of lower metallicity gas seen in merging pairs \citep{Ellison2013}.  
Finally, the creation of classical bulges through mergers is consistent with the fact that both classical bulges and elliptical galaxies follow the same black hole-mass vs bulge velocity dispersion relation while pseudobulges do not \citep{Kormendy2011,Kormendy2013}.
In this model, both classical bulges and supermassive black holes grow through mergers in a process regulated by AGN feedback.

In summary, if classical bulges are formed like elliptical galaxies, it is more likely that the similarities arise due to dissipation in mergers rather than violent relaxation of existing stars. Certainly redistribution of the stars can occur, but may lead to predominantly populating the inner stellar halo \citep{Zolotov2009, Purcell2010} rather than a more central, bulge-like concentration.

 \subsection{Pseudobulges}
The non-spheroidal-like properties of pseudobulges have led astronomers to question their formation through mergers.
As summarized in \citet{Kormendy1993}, initially the evidence for non-merger driven growth came from the disky kinematics of many bulges.
Additionally, mergers tend to drive S\'ersic indices up \citep{VanAlbada1982, Aguerri2001, Kormendy2008}, likely placing them above the normal range for both types of pseudobulges.
More recently, disky pseudobulges have also been shown to deviate from the the photometric projections of the fundamental plane \citep{KK04,Fisher2008}, implying a separate formation path than either elliptical galaxies or classical bulges. 

It is believed that disky bulges grow secularly through the inward transport of material. 
Bars redistribute gas through galaxies by transferring angular momentum to the outer disk.
Near the co-rotation radius gas is collected into rings while gas on smaller orbits is funneled toward the very center \citep[][and references therein]{KK04}.
Furthermore, bars cause gas to shock \citep[see][for a detailed analysis; the concept was first proposed by Prendergast, unpublished c1962]{Athanassoula1992}, resulting in additional inflow. Similarly, other non-axisymmetries like spiral arms can also spur inflow by causing gas to shock.
This fresh supply of gas then enables the formation of a disky psuedobulge.
Observational evidence for gas being funneled to the centers to galaxies comes from the dips in the stellar velocity dispersion at the centers of galaxies \citep{Emsellem2001,Marquez2003,FalconBarroso2006,Peletier2007}.
These regions of low dispersion are likely small central stellar disks formed from the inflow of cold gas.
The observed correlation between bulge and disk scale length adds further support for bulge growth through secular evolution \citep{Courteau1996, Aguerri2005, Carollo2007}.

Similar evidence has been used to argue that B/P bulges from through secular processes.
For instance, there is close correspondence between the existence of B/P shaped bulges and rings or bars \citep{Kuijken1995,Bureau1999} and mergers of disks are unlikely to redistribute stars into the boxy orbits \citep{Bureau1998} characteristic of B/P bulges.  Bars can heat themselves vertically through buckling and resonant star scattering, as has been well established through theoretical modeling \citep[e.g.][]{Combes1981,Combes1990, Pfenniger1990,Raha1991}. 
This is likely the scenario that leads to B/P bulges, so that B/P bulges tend to form in place rather than through transport of fresh material from the outer disk, as is the case for disky pseudobulges.

\citet{KK04} outline three reasons why secular evolution is more likely to create pseudobulges than mergers.  First and foremost is the fact that the most obvious pseudobulges are in barred or oval galaxies, i.e., they are associated with galaxies that have a clear non-axisymmetry that should funnel gas to the center.  Second, they argue that most pseudobulge galaxies show no sign of tidal interactions in progress.  Indeed, time scales may argue against pseudobulges being created in major mergers at high $z$, since many of them are associated with recent star formation and have blue colors.  Finally, \citet{KK04} note that mergers should also heat the thin disk.  Certainly if pseudobulges are created in major mergers at high $z$, they may be correlated with thick disk formation instead \citep{Brook2004}.  However, more recent minor mergers may also lead to the same driving of gas to the central region, without destroying or heating the thin disk as much as previously believed \citep{hopkins09, Moster2010, Moster2012}.  Despite these arguments for inflow of gas to create pseudobulges, other studies \citep[e.g.,][]{Laurikainen2014a} have found that much of the mass in pseudobulges resides in a boxy/peanut/barlens bulge (as opposed to disky pseudobulges), suggesting local heating rather than inflow may dominate the majority of pseudobulge creation.

\section{Theoretical Challenges}
\label{sec:2}

The sheer ubiquity of mergers within a $\Lambda$CDM cosmology allows bulge formation to occur easily in mergers. 
As dark matter halo growth in this cosmology happens through the hierarchical build-up of structure \citep{WhiteANDReese78}, galaxy mergers, especially at high $z$, are predicted to be common.
Major mergers (mass ratio  $\lesssim$1:4) below $z$ = 1 are thought to have occurred in only about half the galaxies \citep{Maller2006}.
However, the increasing merger rate with redshift \citep{Fakhouri2008, Genel2009, Fakhouri10} leads to even higher rates at earlier times. Minor mergers are more common, and we discuss below whether they may be able to induce bulge growth through tidal torques. These theoretical merger rates have been shown to be in agreement with the observed number density of close pairs \citep{Lotz2011} and kinematically disturbed galaxies \citep{Puech2012}.

\subsection{Populations Studies Through Semi-Analytic Models}
Semi-analytic models (SAMs) consist of a set of analytic models applied to a population of halos generated from cosmological simulations that follow only the dark matter.  The dark matter simulation provides the properties of the dark matter halos as a function of time, including their merger history.  The analytic models, in turn, describe the evolution of the baryonic component of the galaxies.  SAMs have used the frequency of mergers to assess their ability to create bulges through the redistribution of stars  \citep[e.g.,][]{Kauffmann1993,Cole1994,Somerville1999}.
Note, though, that SAMs do not distinguish between morphological types of bulges.  They simply lump all bulges together under the label ``spheroid.'' 
In SAMs, mergers above a given mass ratio result in the galaxy's transformation into a spheroid.
Subsequent accretion of gas allows for the formation of a new disk around the spheroid and the galaxy morphology is defined by the spheroid to disk mass ratio.
Explicitly linking bulge formation to major mergers enables SAMs to generally reproduce the observed relationship between morphological type and color \citep{Baugh1996}, the color-magnitude relation \citep{Somerville1999} and the observed morphological mix of galaxies \citep{Cole00}, as well as their approximate environmental dependence.  
Importantly, a merger-driven scenario for spheroids allowed SAMs to reproduce the number density of spheroids at $z = 0$ for galaxies roughly more luminous than $L^{\star}$ \citep{Somerville2014}.  
However, for galaxies fainter than $L^{\star}$, the ubiquity of mergers leads to an overproduction of bulges in low mass galaxies.

\subsubsection{Synergy with idealized simulations}

The analytical prescriptions used to model the baryonic component in SAMs are often shaped by the results of idealized simulations.  Idealized simulations first construct a galaxy according to the simulator's desires, and then follow its evolution.  In idealized binary merger simulations, two such galaxies are built and then allowed to merge under gravity.  Idealized simulations differ from cosmological simulations in that they are computationally much less expensive, since they neglect the cosmological context.   They generally do not include subsequent gas accretion onto the galaxies, or torques from large-scale structure. This allows idealized simulations to reach relatively high resolutions while exploring a large range of parameter space more quickly than cosmological models.  This ability to explore parameter space makes them ideal to derive analytical prescriptions that can be fed into SAMs.

The overproduction of spheroids in fainter galaxies in SAMs can be alleviated when the gas fractions of merging galaxies are considered.  
As shown in the binary merger simulations of \citet{Hopkins2009a}, angular momentum loss from gas primarily occurs through internal torques generated by the merger.  
In these situations, the dissipational gas bar will lead the dissipationless stellar bar in phase.
The resulting gravitational torque causes angular momentum to be transferred from the gas to the stars \citep{Barnes1991}.
In gas rich mergers, the relatively low mass of the stellar bar will result in less angular-momentum loss and lower bulge masses.  
This scenario has been further backed-up by simulations of gas-rich disk mergers \citep{Springel2005, Governato2009} and observations that find a significant fraction of $ z \sim$ 1 galaxies had sufficient gas fractions to rebuild their disk \citep{Hammer2009}.
This dependency on gas fraction results in less efficient bulge formation in lower mass galaxies \citep{Hopkins2010a} and high $z$ galaxies \citep{Stewart2009}.  
In cosmological simulations, the trend with gas fraction ultimately makes the final bulge-to-disk ratio highly dependent on the ability of stellar feedback to limit star formation prior to and during the merger \citep{Robertson06}.

While including the role of gas in the resulting bulge-to-disk ratios brings the number of spheroids in low mass galaxies into agreement with observations \citep{Hopkins2009a,Porter2014}, it may also lead to an under-prediction of the number of spheroids at the massive end.  A number of SAMs have found that an additional spheroid creation mechanism, such as those resulting from disk instability, may be required to create spheroids in galaxies brighter than $L^{\star}$\citep{Parry2009, DeLucia2011, Porter2014,Brennan2015, Somerville2014}.

\subsection{The Problem of pseudobulges}
While the SAMs may be able to explain the frequency of spheroids in $L^{\star}$ galaxies and fainter, recall that they do not distinguish between classical and pseudobulges.  
The frequency of pseudobulges (and apparent lack of classical bulges) in galaxies as massive as the Milky Way raises a potential challenge for $\Lambda$CDM \citep[e.g.][]{Weinzirl2009,Kormendy2010,Peebles2010}.
Hierarchical growth is such that interactions and mergers are common at every redshift for massive galaxies. 
For example, \citet{Stewart2008} found that the majority (70\%) of Milky Way-mass halos have experienced at least one merger with an object 10\% the mass of the Milk Way or greater.
If mergers lead to the formation of classical bulges rather than pseudobulges, then it is very mysterious that classical bulges are not more common.  
Indeed, it is puzzling that classical bulges are not the {\it dominant} form of bulge found in the Local Universe.

To reconcile the observed frequency of pseudobulges with the hierarchical nature of $\Lambda$CDM, one or more of the following possibilities must be true: 1) the merger rates must have been over estimated due to errors in converting halo merger rates to stellar galaxy mergers, 2) pseudobulges must be able to camouflage already-existing classical bulges, 3) pseudobulges can also be formed in mergers, or 4) classical bulge formation during mergers must be less efficient than currently thought, possibly because of feedback or because mergers are happening between galaxies with higher gas fractions.  

As far as the first possible solution is concerned, while there is some uncertainty in the merger rates of galaxy disks, {\em halo} merger rates are very well established and significantly reducing galaxy merger rates would likely require assuming a different matter power spectrum on small scales than generally adopted in $\Lambda$CDM models.

In regards to the second possible solution, the existence of bulges containing both pseudo and classical bulge characteristics, i.e. composite bulges, can be presumed from the overlap in populations \citep{Gadotti2009}. A number of individual galaxies with composite bulges have been identified \citep{Nowak2010, Mendez-Abreu2014, Erwin2014}. 
Many of theses galaxies were previously thought to contain only a classical or pseudobulge, so it is clear that either type of bulge can dominate and mask the presence of the other.
Unfortunately, since 1-D surface-brightness profiles cannot clearly distinguish between pure pseudobulges and composite bulges \citep{Fisher10, Erwin2014}, and because it is difficult to obtain stellar kinematic information for a wide sample of galaxies, it is currently not possible to make firm estimates on the fraction of pseudobulge-identified galaxies that also contain classical bulges.
\citet{Mendez-Abreu2014} found composite bulges in 70\% of their sample of barred galaxies; these composite systems generally had large photometrically-defined classical bulges with an inner pseudobulge-like structure.
\citet{Erwin2014} in turn estimated that at least 10\% of S0--Sb barred galaxies were composite-bulge systems and indeed had difficulty entirely ruling out the presence of classical bulges in any of the pseudobulge-identified systems they looked at.
We can therefore conclude that the observed frequency of classical bulges may have been underestimated.
 However, galaxies in which classical bulges are obscured by pseudobulges do not appear common enough to account for all of the discrepancy between observation and theory.
 Furthermore, the small classical bulges that are masked by pseudobulges in observations tend to contain a much lower stellar fraction than the classical bulges produced by mergers in simulations, as we discuss in the next section.
We must, therefore, turn to the third and fourth possible solutions: pseudobulge formation during mergers and less efficient classical bulge formation during mergers.
We examine the latter in detail in the following sections but for now, we concentrate on the possibility that some pseudobulge growth could be merger-induced.

The presumed formation of pseudobulges through non-axisymmetries raises some potential issues with distinguishing secular from merger-driven bulge growth.
Minor mergers and interactions with satellites can in some cases induce bar formation and result in pseudobulges \citep[e.g.,][]{Eliche-Moral2006, Guedes2013}.
Should bulges formed in this manner be considered the result of secular evolution?
Alternatively, what do the bulges look like that are created when gas funnels to the center of a merger remnant and forms stars?  Would they look like the dissipative collapse that can form spheroidals with large $n$ (discussed in Section 2.1.1), or would the fact that the gas is funneled internally from the galaxy result in a bulge that had the tell-tale signs of disky pseudobulges?  

Idealized binary merger simulations should be able to address these questions. Unfortunately, one must be very careful to understand resolution effects.  
The merger simulations of \citet{Keselman2012} adopted 70 pc force resolution for newly formed stars. They examined the resulting bulge both through a decomposition of the projected density profiles, and kinematically.  In all cases, the central component had $n < 1$, and in almost all cases this component is also rotationally supported (at least for a few Gyr after the merger).  Hence, for the first time, these authors have demonstrated that mergers of galaxies appear to produce pseudobulges rather than classical bulges.  

A similar result was hinted at in the study of ``extra light'' as observed in the central surface brightness profiles of some ellipticals. 
This extra light is though to be the signature of a central starburst produced by gas that is funneled to the center during mergers \citep{Kormendy1999, Hopkins2008a}, much like the process expected to create bulges in disk galaxies in mergers.  
The extra light components could be fit with $n \sim 1$, but with a caveat: the extra light component \citep[and presumably the bulges in ][]{Hopkins2009a} were comparable in size to the force resolution of the simulations.   
Resolution effects would act to artificially flatten the central profile, and reduce the $n$ of the central light component. 

In principle, any computational study of the $n$ value of the resulting bulge formed in mergers needs to undergo a rigorous convergence test.
These two simulations, therefore, present promising results that need to be replicated in simulations that can resolve the bulge structure.
Once the bulge can be resolved, it will be possible to check whether mergers can produce, in addition to the lower $n$  and the rotational kinematics already hinted at, the other characteristics typical of disky pseudobulges: recent or on-going star formation, dense gas, and morphological substructures such as bars, rings, and spiral arms.
It is especially important to examine these other characteristics since so much of the evidence for pseudobulge growth being secularly driven is based on morphological substructures, for instance the strong correlation between pseudobulges and the presence of bars and ovals.
Additionally, the ubiquity of star formation within pseudobulges implies that pseudobulge growth is not dominated by episodic events, such as mergers.

In summary, the idea that mergers always lead to classical bulge formation in disk-dominated galaxies is problematic within a hierarchical model like CDM.  Historically, the ubiquity of mergers led to an overproduction of bulges in $L^{\star}$ galaxies and fainter.  The formation of bulges was shown to be suppressed in gas-rich mergers, and the adoption of a model that accounted for gas fraction in bulge formation can reproduce the observed trend in morphology with mass up to $L^{\star}$ \citep{Hopkins2009a}.  However, if the remaining bulges that do form in mergers are classical, then tension still remains with CDM theory given that the Local Volume is dominated by pseudobulges \citep{Kormendy2010, Fisher2011}.  One possible solution to alleviate the tension is if some pseudobulges are induced by minor mergers that trigger bar formation.  Indeed, preliminary work using idealized merger simulations suggests that pseudobulges may be the outcome of gas inflow that leads to centralized star formation in gas-rich disk galaxy mergers \citep{Keselman2012}.  However, a counter-argument to this is that fewer galaxies show rings, ovals, or bars than are likely to have recently experienced minor mergers, suggesting that minor mergers cannot always form pseudobulges, if they ever do.

If disky pseudobulges are not formed during mergers, then it is left to other secular processes and disk instabilities to form them.  
Given the dearth of classical bulges and evidence that many pseudobulges are boxy that likely formed in situ due to heating from a bar \citep{Laurikainen2014a}, it is likely that secular evolution is the primary creator of disky pseudobulges.  
However, this still requires that the formation of bulges (of all types) must be suppressed in mergers in order to match their low numbers in the local Universe.  We explore this possibility below.

\section{Forming Bulges in Cosmological Simulations}
\label{sec:3}
Cosmological simulations are the ideal place to test the physics of bulge formation, owing to the fact that the complex processes of mergers, stellar feedback, and gas inflow and outflow can be self-consistently modeled.  
Until recently, however, it has been very difficult to specifically study bulge formation because bulges are generally on the same size scale as the force resolution convergence.  
Convergence in the density profile is only achieved when enough particles are enclosed that the time scale for collisional relaxation of the particles is longer than the age of the Universe \citep{Power2003}.  In practice, this means that densities usually converge at 4--6 times the force softening length \citep{navarro10}.  The highest resolution simulations yet of Milky Way-mass galaxies have achieved a force resolution of $\sim$70 pc \citep{Hopkins2013d}, but most simulations published to date have had force resolutions on the order of $\sim$150 pc or more, making the regions interior to $\sim$700 pc officially unresolved.  
As most bulges have effective radii between 100 and 1000 pc \citep{Fisher10}, these resolutions are insufficient to study the structure of the majority of bulges.
However, additional simulations with sub-100 pc force resolutions will be achieved within the next few years, and thus the ability to begin resolving bulges and analyzing their growth is becoming a realistic possibility.

Historically, simulations have tended to produce galaxies that are too compact, with a large central mass concentration and large stellar spheroid \citep[e.g.,][]{sn99,ns00, Abadi2003c, governato04, cecilia2}.  The resulting overly-massive and concentrated bulges are the direct result of the ``overcooling'' problem in simulations.
In this section we discuss the origin of overcooling, and describe the recent successes in feedback modeling that appear to overcome the overcooling problem.  In theory, a fully successful model must also match the observed stellar mass--halo mass relation and its evolution in order to ensure that gas is not being over-consumed in star formation.  Over-consumption leads to lower gas fractions, and drives the building of larger bulges than observed. Unfortunately, no simulation that has demonstrated its ability to match this relation back to high $z$ has yet been used to study bulge formation.  We instead highlight a few of the best results to date, and point out the open questions that future high resolution cosmological simulations can address.

\subsection{The Problem of Overcooling}

In the overcooling problem, baryons at early times cool rapidly to the center of halos, resulting in galaxies with dense concentrations of gas and stars. In a hierarchical formation model like CDM, these galaxies undergo multiple subsequent mergers.  During these mergers, orbital angular momentum is transferred to the dark matter of the accreting halo through dynamical friction.  By the time the dense baryons arrive at the center of the accreting halo, little angular momentum remains in them and the resulting galaxies show the classic signs of the angular momentum catastrophe \citep{Navarro1994,Katz1994,maller02b, donghia06}.

Including appropriate stellar feedback has been the most effective method for reducing overcooling in simulations \citep{Governato2007, cecilia1, piontek09a}.
Feedback defends against early cooling of gas in simulated halos by creating a hot gas reservoir that only allows gas to cool onto galaxies at later times.  Ideally the feedback hinders cooling prior to the period of rapid mergers, which prevents angular momentum loss in tidal effects and leads to more realistic mass distributions in galaxies \citep{Robertson04,Okamoto05,cecilia1, zavala08, keres09b, piontek09a}.  Stellar and supernova feedback in the disk seems to be the most important source of heating \citep[e.g.,][]{Governato2010, Agertz2013, Aumer2013, Ceverino2013, Hopkins2013d}, though pre-heating at reionization also prevents gas from cooling in the lowest mass halos \citep{Quinn96,thoul, gnedin00,Okamoto2008}.
As we will discuss later, stellar feedback has other positive effects, such as maintaining the gas fractions of disks and expelling low angular momentum baryons.

\subsection{Feedback Implementations and their Effects on Central Concentration}
Stellar feedback offers a promising avenue for reducing the mass and concentration of bulges.
However, simulators must grapple with the fact that both star formation and feedback take place on scales much too small to be resolved in cosmological simulations.
For example, simulations must adopt a prescription that mimics star formation on kiloparsec scales, rather than the sub-parsec scales where star formation actually takes place (the scheme is then termed ``sub-grid'').  
Thankfully, star formation on galaxy scales does appear to follow a global trend, the Kennicutt-Schmidt relation \citep{Kennicutt98, MK01}.  However, even if modelers can determine the star formation rate based on the Kennicutt-Schmidt relation, this does not guarantee that the feedback scheme used for subsequent supernova feedback will satisfy observed constraints such as the Tully-Fisher relation or the mass-metallicity relation.  

Early simulations found that supernova energy was quickly radiated away in the surrounding dense medium without impacting the galaxy \citep[particularly at high $z$ when the gas is more dense, e.g.,][]{katz92, sn99}.  
Simulators have thus been forced to develop sub-grid recipes for feedback to model how the energy is transferred to the interstellar media.
One way to avoid the immediate radiating away of supernova energy is to turn off cooling in gas particles within the ``blastwave'' of the supernova remnant for a period of time \citep{MO77, thacker00b, Stinson06}.  While this cooling delay is designed to mimic the sub-resolution adiabatic expansion of the supernova, it is often considered undesirably artificial.  Others avoid disabling cooling by adopting a ``multiphase'' gas particle model that prevents hot gas particles from being artificially influenced by their cold gas nearest neighbors \citep{Hultman1999a, marri03, Harfst2006a}.  However, the disconnect between cold and hot gas in this approach can also be considered unphysical and problematic. 

More recent works have instead embraced additional sources of energy from young stars \citep{Hopkins2011d, Agertz2013, Aumer2013, Kannan2013a, Trujillo-Gomez2013, Sales2013, Stinson13, Wise2012} and in these schemes of constant energy injection, cooling need not be turned off for feedback to have a strong effect.
Essentially, supernova energy can more easily escape in these schemes because in the $\sim$4 Myr prior to any type II supernova, the massive stars have already contributed to the destruction of their natal birth clouds \citep{Agertz2013}.  There are multiple early energy sources which may contribute, e.g., photoionization of the clouds via UV radiation, momentum injection from stellar winds, and radiation pressure on surrounding dust grains \citep[e.g.,][]{Murray2005, Sharma2011, Murray2011a, Zhang2012b, Lopez2013}, but many of them are still poorly constrained.  In particular, the strength of the radiation pressure is highly debated.  While it is possible that IR trapping causes a single photon to bounce a number of times, increasing its effectiveness, it has also been suggested that a full model of radiative coupling will produce chimneys by which the photons escape, making them ineffective \citep{Krumholz2012, Krumholz2013}.  Despite the debate over the details, the overall trend is to input more energy into the ISM, which has allowed cosmological simulations to successfully reproduce a number of observed galaxy scaling relations in recent years \citep[e.g.,][]{Brook2012d, Aumer2013}.\footnote{\citet{Agertz2010} is notable for producing a galaxy with a relatively low bulge-to-total ratio using low star formation efficiencies, rather than high levels of feedback (B/T = 0.21 for a 1.25 $\times 10^{12} \Msun$ halo with their preferred feedback model).  However, this galaxy produced too many stars overall. As discussed in \citet{Agertz2013}, stellar feedback is necessary for producing galaxies with {\em both} appropriately low stellar masses and small bulges.}

Along with the increased energy deposition into the ISM, increasing computational power has allowed a change to the distribution of star formation in simulations.  Cosmological simulators are now resolving gravitational forces on scales $<$ 100 pc.  These high resolutions allow high density peaks in the gas to be resolved.  When these peaks are not resolved, star formation takes place in diffuse, warm gas across the entire disk and the supernova energy is quickly radiated away \citep{ceverino09, saitoh08}.  
On the other hand, if high density peaks are resolved and and the star formation is limited to these peaks (and gas is allowed to cool below 1000K), the supernova energy is concentrated into smaller physical regions.  The overall result is that feedback creates over-pressurized regions and becomes more effective, as well as highly localized.  Localized feedback naturally drives gas outflows \citep[galactic ``winds,''][]{Governato2010, Guedes2011, Christensen12a, Hopkins2013d, Agertz2014a} without the need to implement a separate numerical prescription for outflows \citep{Dave2011, Marinacci2013, Vogelsberger2013}.  Galactic winds appear to be ubiquitous \citep[e.g.,][]{martin05, vcb05, weiner08}, so the natural driving of outflows in high resolution simulations can be considered a major success.

These high resolutions with galactic winds led to the simulation of realistic central mass distributions in low mass dwarf galaxies for the first time: bulgeless dwarf galaxies \citep{Governato2010, Governato2012, Teyssier2013, DiCintio2013}. Bulgeless disks (no classical or pseudobulge) are more common in low mass dwarf galaxies \citep{dutton09,Kormendy2014}. 
In these galaxies, high resolution and localized feedback lead to gas outflows that preferentially remove low angular momentum gas \citep{Brook2011a, Brook2012c}.
If retained, this low angular momentum gas could have formed a large stellar bulge.
Instead, the loss of it results in a stellar disk with a purely exponential surface brightness profile and higher specific angular momentum than predicted otherwise, consistent with observations \citep{Donghia2004, Dutton2009a, Governato2010}. 

Ideally, the same star formation and feedback prescription that produces realistic bulgeless dwarf disks would form realistic bulges as halo mass increases.  One might imagine a scenario in which the deeper potential wells in high mass galaxies prevent the complete loss of low angular momentum material \citep{Dutton2012} so that these galaxies still form bulges.  In this scenario, outflows are still required to remove some low angular momentum material at high $z$ in order to match observed bulge sizes even in Milky Way-mass galaxies  \citep{Binney2001a, bullock01, VandenBosch2001c, VandenBosch2001a, VandenBosch2002}.  
If less low angular momentum material is lost as galaxy mass increases, this trend could reproduce bulge prominence as a function of galaxy mass. 

In practice, it is difficult to reach the same high resolution currently achieved in dwarf galaxy simulations in more massive galaxies.  Being more rare, massive disk galaxies like the Milky Way require a larger simulation volume.
A larger volume is also necessary to ensure that the large scale tidal torques that deliver angular momentum to the galaxy are included \citep{white84, barnes87}.  To reach sub-100pc resolutions requires significantly more particles, making this currently computationally challenging, although feasible for the first time \citep[e.g.,][]{Hopkins2013d}.  No works have yet examined the bulge properties of massive disk galaxies formed at these high resolutions, though.

\subsection{Bulge Formation Studies with Cosmological Simulations}

The highest resolution studies to yet examine bulges in massive disks in fully cosmological simulations have $\sim$150pc resolution \citep{Guedes2011, Christensen14}.  Most works have examined central mass distributions by restricting themselves to a study of circular velocities ($v_c$) and the bulge-to-total (B/T) ratios of their galaxies \citep[e.g.,][]{Scannapieco2010, Stinson13, Aumer2013}.  The former is useful because large bulges lead to a declining rotation curve rather than the flat curves observed in galaxies.  The latter offers many opportunities to compare with observations.  A comparison of B/T ratios must be done carefully, though.
A number of works have noted that the bulge fractions determined through kinematic decomposition are systematically larger than those determined through photometric decomposition \citep{Governato2009, Scannapieco2010, Marinacci2013, Aumer2014, Christensen14}, although the extent of the discrepancy varies widely with simulation and decomposition methodology.  
For instance, in the papers cited above, kinematic decomposition produces bulge mass anywhere from 1.25 to 12 times greater than the photometrically determined mass.
The need for photometric decompositions (on simulated observations in the appropriate bands and using standard observational decompositions methods) when comparing with observed bulge properties is clear.
However, even in the most extreme cases, using photometric rather than kinematic decomposition does not eliminate the problem of too massive or concentrated of bulges.

A few modelers have investigated the structure and growth of their bulges in detail.  \citet{Guedes2013} and \citet{Okamoto2013} both found that their bulges were best fit by a S\'ersic index $<$ 2 (1.4 in the former case, 1.4 and 1.2 in the latter case, as measured through $i$-band photometric decomposition), leading these authors to classify them as pseudobulges (again, we note that resolution may artificially lower $n$, as these simulations would not necessarily yield converged results on scales smaller than $\sim$750 pc).  
Despite being classified as pseudobulges, their formation is inconsistent with slow growth via secular evolution processes.  Instead, these bulges form fast at high redshift, and mostly in situ from gas funneled to the center of the main galaxy rather than from accreted material.  
\citet{Guedes2013} noted that mergers contributed a similar fraction of accreted stars to both the bulge and the high $z$ disk, rather than preferentially to the bulge and, according to \citet{Pillepich2014b}, only a quarter of the final bulge stellar mass was formed ex situ.
In two of the three galaxies examined between \citet{Guedes2013} and \citet{Okamoto2013}, the presence of a bar was tied to bulge formation.  
However, \citet{Guedes2013} found that the bar formation was itself induced by tidal interactions with accreting galaxies.  Hence, mergers were indirectly responsible for the growth of a bulge with a low S\'ersic index in their simulation.  

Unfortunately, it is unclear how much merger history plays into the fast pseudobulge formation that was found in these simulations. Most of the galaxies studied so far were chosen to have quiescent merger histories, with their last major mergers occurring at $z \gtrsim 1$.  Are the early formation times of these simulated bulges due to a biased selection with all major mergers occurring at high $z$?  Would the bulges continue to grow, or become more classical, with lower $z$ mergers?   

One other important consideration is that both \citet{Guedes2013} and \citet{Okamoto2013} suffer from too much star formation at high $z$ compared to abundance matching results \citep[see also][]{Stinson13, Agertz2013}.  Forming too many stars at high $z$ suggests an overconsumption of gas -- and hence lower gas fractions than are realistic.  If the simulated galaxies are gas-poor galaxies at high $z$, then their merger-driven bulge growth would be too efficient \citep{hopkins09, Hopkins2009a}.  Higher, more realistic gas fractions could possibly be achieved by increased amounts of feedback.
Indeed, both \citet{Guedes2013} and \citet{Okamoto2013} conclude that more feedback at high $z$ (possibly in the form of AGN feedback) is likely necessary to form the large, bulgeless disk galaxies that are found in the Local Volume \citep{Kormendy2010, Shen2010, Fisher2011, Laurikainen2014a}.

Recently, \citet{Aumer2014} analyzed the evolution of both the disk and bulge in simulations with greater amounts of feedback. Unlike \citet{Guedes2013} and \citet{Okamoto2013}, these simulations included additional feedback from young stars.
These simulations did indeed produce more realistic star formation histories for their stellar mass \citep{Aumer2013} and had low $i$-band B/T values (typically less than 0.15).
However, the stellar distributions reveal discrepancy with observations.  Specifically, the simulated Milky Way-mass galaxies show both too little central growth combined with too much outer disk growth compared to results in \citet{VanDokkum2013}.  
It would be tempting to conclude that these unrealistic stellar distributions were the result of ejected bulge gas being recycled to the outer disk.
However, \citet{Ubler2014b} studied these same simulations and showed that most recycled gas returned with a similar angular momentum as when it was ejected.
\citet{Aumer2014} conclude that their feedback appears to be too strong at $z < 1$, while providing more accurate star formation at higher $z$.  Young star feedback has also been shown to produce disks that are too thick compared to observations \citep{Roskar2013}.  Taken together, these results suggest that more feedback is necessary at high $z$ to match star formation rates, but that current models of feedback from young stars (prior to supernovae) creates new problems.  Could the addition of AGN feedback at high $z$ reduce the need for such strong stellar feedback?  Or do we simply have yet to understand how stellar feedback operates at high $z$?

Overall, spiral galaxies with appropriate (if slightly on the massive end of the observable range) bulge-to-total ratios have now been produced and analyzed in a handful of simulations \citep{Guedes2013, Okamoto2013, Aumer2014}.
However, as of yet there has been little investigation into how realistic the other bulge properties are.
While \citet{Christensen14} showed simulations that fit observed bulge scaling relations, those bulges were overly massive compared to the disks, having $H$-band B/T values of 0.43 and 0.53.
Similar analysis of the scale lengths, surface brightnesses, and colors of simulated bulges is needed in future simulations, in addition to the more common bulge-to-total ratio.

Moreover, most identification of simulated bulges as being classical bulges or pseudobulges has been based entirely on the S\'ersic index.  
While lower S\'ersic indices are well correlated with the disky-properties that indicate a pseudobulge in observed galaxies, it is not clear that simulated galaxies yet produce as clean of a divide.
Until it has been shown that pseudo and classical bulges in simulations have the same sets of identifying features as observed galaxies, more holistic classification of bulges in simulations is necessary.
Most importantly, simulated bulges should have the defining feature of pseudobulges: being disk-like in terms of a flatten morphology and rotation-dominated stellar kinematics.
They should also reproduce the observed sub-types of pseudobulges: disky versus B/P.
For instance, simulated bulges should frequently include morphological features such as bars, rings and spiral arms, recent or on-going star formation, and an abundance of dense gas as these are typical observed characteristics of disky pseudobulges.  

Reproducing both types of pseudobulges in cosmological simulations is highly dependent on whether the simulations can resolve the instabilities that create bars and other asymmetries.  B/P bulges are tied to the presence of bars and it is not clear whether cosmological simulations capture bar formation and destruction.  More specifically, long bars can and do form when simulations resolve the perturbation scales sufficiently, but the simulations may not capture the smaller scale disturbances that form shorter bars.  Once a bar does form, it is not clear if the simulations will then capture the processes that should lead to its destruction.  However, as cosmological simulations achieve ever higher resolution, they are approaching the resolution scales that isolated disk galaxy studies have used to study similar processes.  Hence, the ability to answer the question of whether simulations capture these processes is soon to be within reach.

\section{Limiting merger-driven bulge formation with stellar feedback}
\label{sec:4}

Results from both idealized merger simulations and fully cosmological simulations lead us to conclude that the primary mechanism of bulge growth in mergers is tidal torquing of gas within the galaxy, which drives gas into the central regions to undergo a burst of star formation.  Recent results from the study of bulgeless dwarfs, however, suggests another step in the story.  Bulgeless dwarfs arise in cosmological simulations regardless of their merger history.\footnote{Due to the required high resolution to form bulgeless dwarf galaxies, a large, statistical sample is difficult to produce.  Some observed dwarfs do have bulges, but bulgeless disks dominate the population \citep{dutton09}. The current small numbers of simulated dwarfs has limited our ability to produce dwarfs with bulges to understand their origin.}  While mergers in dwarfs drive gas to the center, subsequent bursts of star formation drive galactic winds that remove that gas from the galaxy.
Essentially, mergers in dwarfs can lead to the removal of bulge material, rather than the creation.  To what extent could this scenario hold in more massive galaxies with deeper potential wells?  At what scale do winds fail to prevent bulge formation?

Some insight into these questions comes from \citet{Brook2012c}, which examined the history of a disk galaxy of 2$\times$10$^{11}$ M$_{\odot}$ in virial mass.  
Although this galaxy did eventually form a bulge, the formation was considerably delayed compared to the more massive disk galaxies discussed in the previous section.  In fact, the galaxy was still bulgeless at $z=1$.  
Bulge formation at higher redshifts was suppressed because it was easier for this galaxy to lose its low angular momentum gas when the potential well of the galaxy was shallower, i.e. earlier in the galaxy's history \citep{Brook2012c}.

Of course, merger rates were also higher early in the history of the Universe.  
Is it possible that the shallower potential wells of galaxies at high redshifts (and therefore the ability of winds to escape) offset the effects of the increased merger rates?
Typically, the existence of massive, bulgeless disk galaxies at $z=0$ in the Local Volume has been viewed as a challenge to CDM.  If mergers instead drive a burst of star formation that expels low angular momentum gas, could the merger rate predicted in CDM instead be viewed as the solution to the existence of massive, bulgeless disks?
In this section we explore this possibility in more detail.

\subsection{Trends with Galaxy Mass}
The mass loading factor (the mass of gas outflowing from the galaxy divided by the mass of stars formed) depends strongly on galaxy mass.
Models typically assume an exponential scaling with circular velocity of either $-1$  for ``momentum driven winds'' or $-2$ for ``energy driven winds."
These scalings imply that dwarf galaxies have mass loading factors a couple of orders of magnitude larger than those of Milky Way mass galaxies.
As such, ten to a hundred times more gas is expelled from the central regions of lower mass galaxies than is formed into stars.
In other words, as galaxy mass increases, we expect a smaller fraction of the material driven to the centers of galaxies through mergers to be expelled.  
This should lead to increasingly dominant bulges with increasing galaxy mass.

Here, we examine the extent to which mergers are able to drive bulge growth as a function of galaxy mass in a current set of high resolution cosmological simulations.
We compare a suite of ten simulated galaxies (Table 1), all selected from ``zoom-in'' galaxy simulations produced with the Smoothed Particle Hydrodynamic code, {\sc gasoline} \citep{gasoline}.
These galaxies cover a mass range of $10^{10}$ to $10^{12} \Msun$ at redshift zero.
They were selected on the basis of having experienced a 10:1 or greater mass ratio merger since $z$ = 3.

The version of {\sc gasoline} used to produce nine of these simulations is described in detail in \citet{Christensen2012}.
In brief, though, the simulations have gas particle masses between 3.3 and 27.0$\times$10$^3 \Msun$ and softening lengths between 87 and 174 pc.
Star formation occurs probabilistically according to the free fall time and the local molecular hydrogen abundance.
Supernova feedback is implemented using a blastwave scheme \citep{Stinson06} with the total amount of energy deposited in the ISM being 10$^{51}$ ergs per supernova.
This version of the code has been successful at reproducing many observed properties of galaxies, including the shapes of bulges \citep{Christensen14}, the cores of dwarf galaxies \citep{Governato2012}, the stellar mass to halo mass relation at $z=0$ \citep{Munshi2013}, gas fractions at $z=0$ \citep{Munshi2013}, and the mass distribution of satellite and field dwarf galaxies \citep{Zolotov2012,BZ14}.

The tenth galaxy is produced by version of {\sc gasoline} with a newer model of supernova feedback, as described in \citet{Keller2014}.
This feedback model replicates the effect of superbubbles on the surrounding gas.
Superbubbles are generated by clustered star formation when the individual winds and supernovae merge.
They are also much more efficient at generating gas motion than individual supernovae.
For instance, \citet{Keller2014} found that in a Milky Way-mass galaxy, they drove ten times more mass in outflows as the blastwave feedback model.

\begin{table*}
\caption{Characteristics of the examined galaxies and their mergers.  Galaxy 5 experienced two qualifying mergers and is, therefore, listed twice. Galaxy 10 is from a separate set of simulations with the more efficient superbubble feedback recipe. The ``Primary Halo Mass" listed in column 3 is the mass of the primary at the start of the listed merger. }  
\centering 
\begin{tabular}{c c c c c c} 
\hline\hline 
Halo & Mass at $z$ = 0 	& Primary Halo Mass &Merger Redshift  &Amount of Time for Merger 	& Merger Ratio\\ 
  	& [$10^{9} \Msun$]	& [$10^{9} \Msun$]	&                             & [Gyr]					&			\\
\hline 
1 	& 23 		& 7.1 	& 1.3 	& 2.1 	& 1.2 \\ 
2 	& 38 		& 28 		& 1.1		& 1.1		& 3.5 \\ 
3 	& 38 		& 2.1 	& 1.1		& 2.6		& 1.5 \\ 
4 	& 43 		& 22	 	& 1.9		& 1.1		& 7.7 \\ 
5 	& 180 	& 19	 	& 1.2		& 2.1		& 2.1 \\ 
`` 	& `` 		& 110	& 1.1		& 0.9		& 9.4 \\ 
6 	& 340 	& 160	& 1.3		& 1.1		& 2.2 \\ 
7 	& 770 	& 360	& 1.6		& 1.3		& 1.1 \\ 
8 	& 880 	& 59		& 1.7		& 2.5		& 1.1 \\ 
9 	& 910 	& 330	& 1.5		& 1.2		& 1.2 \\ 
\hline
10	& 800	&140		& 2.6		& 1.0		& 2.0 \\ %
\hline 
\end{tabular}
\label{table:nonlin} 
\end{table*}

\begin{figure*}
\begin{center}
$
\begin{array}{lr}
\includegraphics[width=0.50\textwidth]{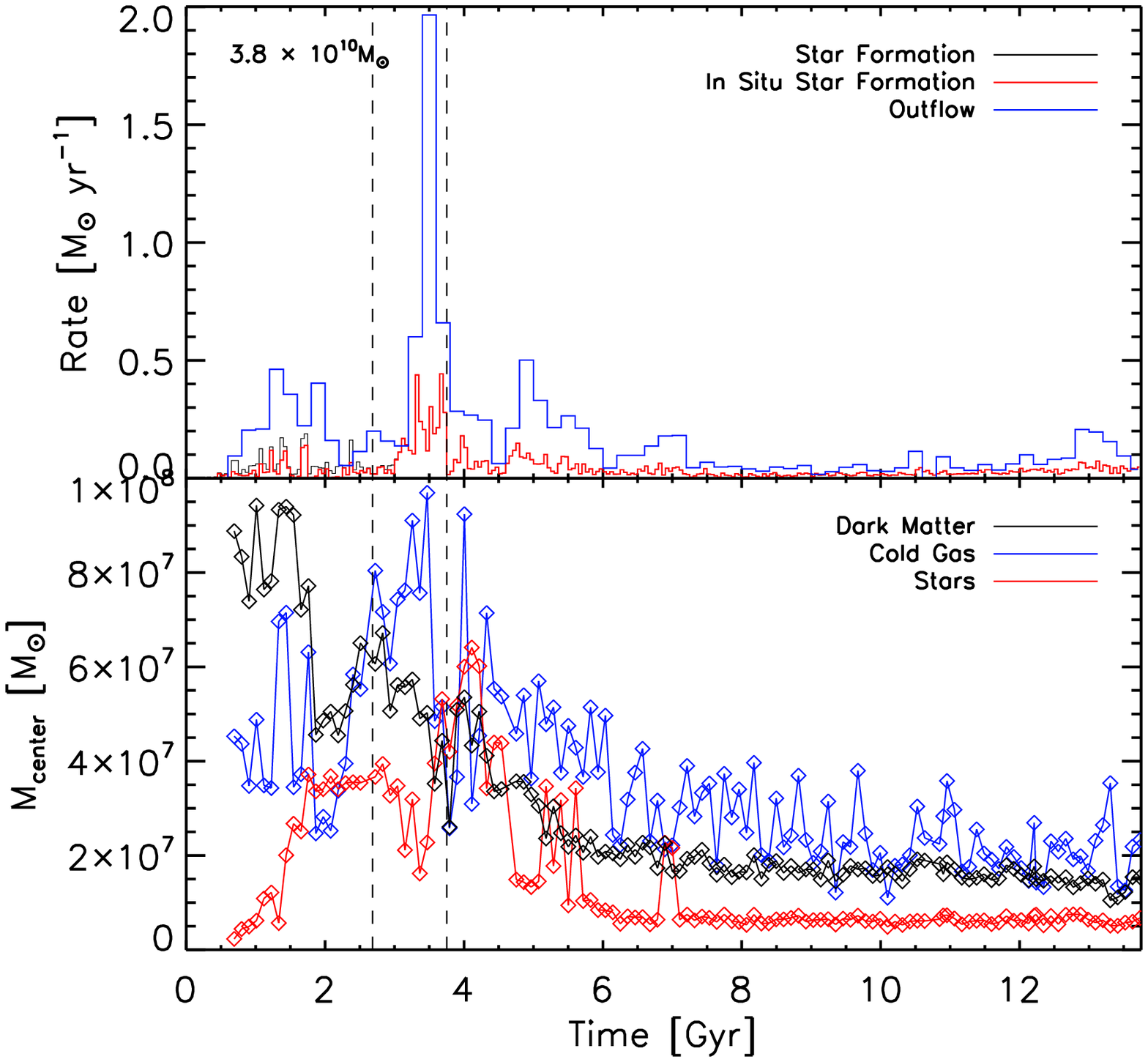} 
\includegraphics[width=0.50\textwidth]{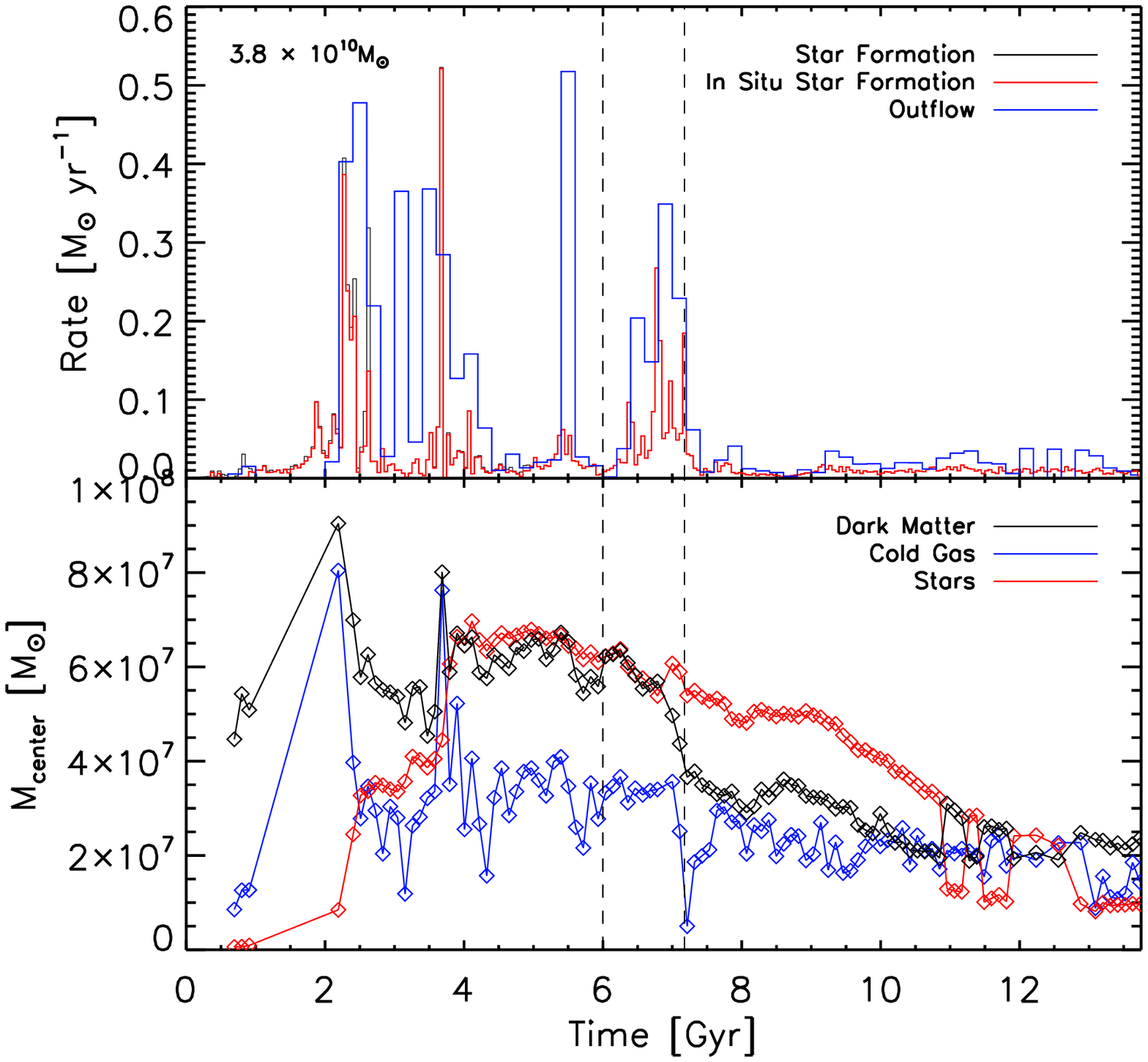}\\ 
\includegraphics[width=0.50\textwidth]{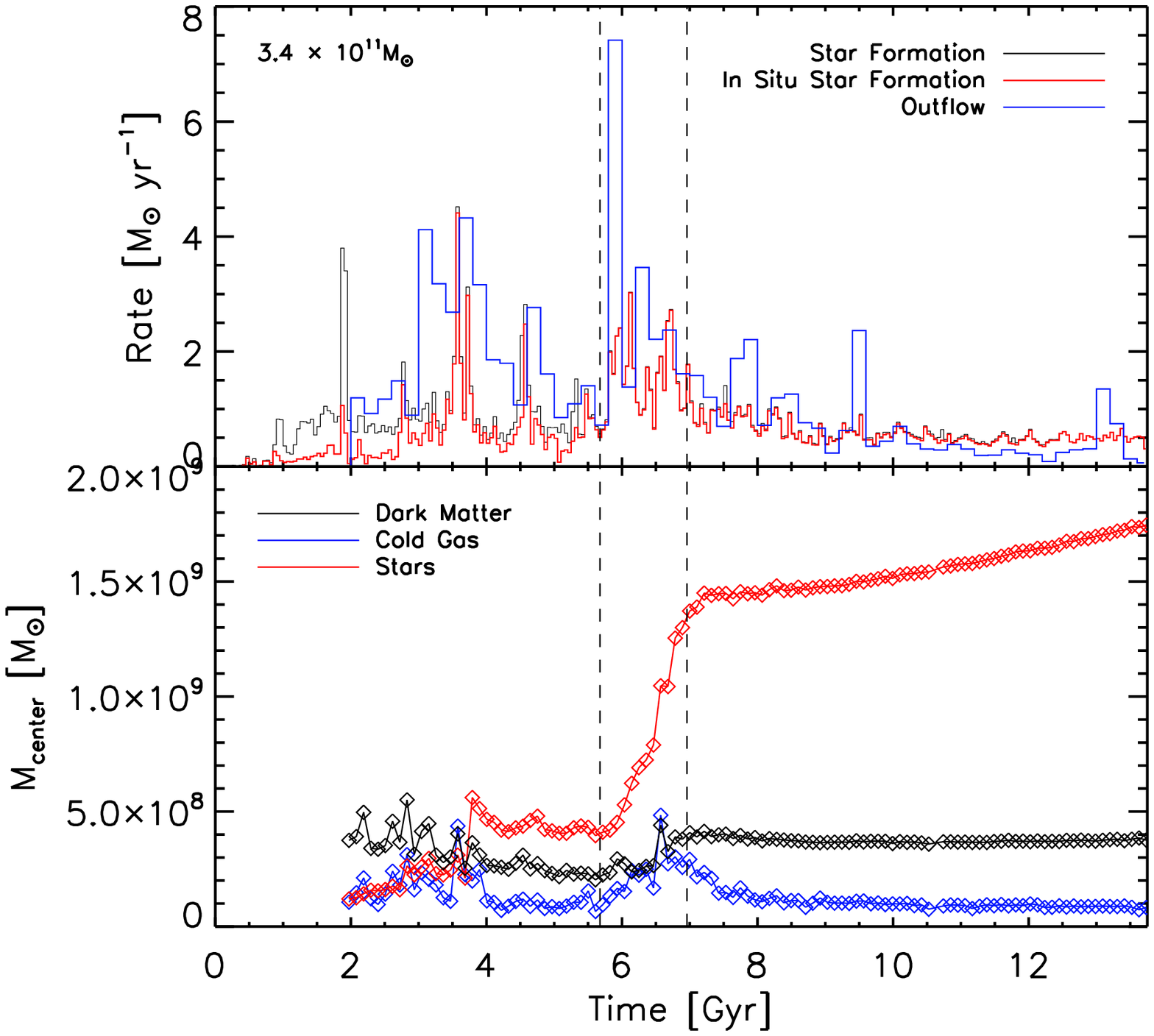}
\includegraphics[width=0.50\textwidth]{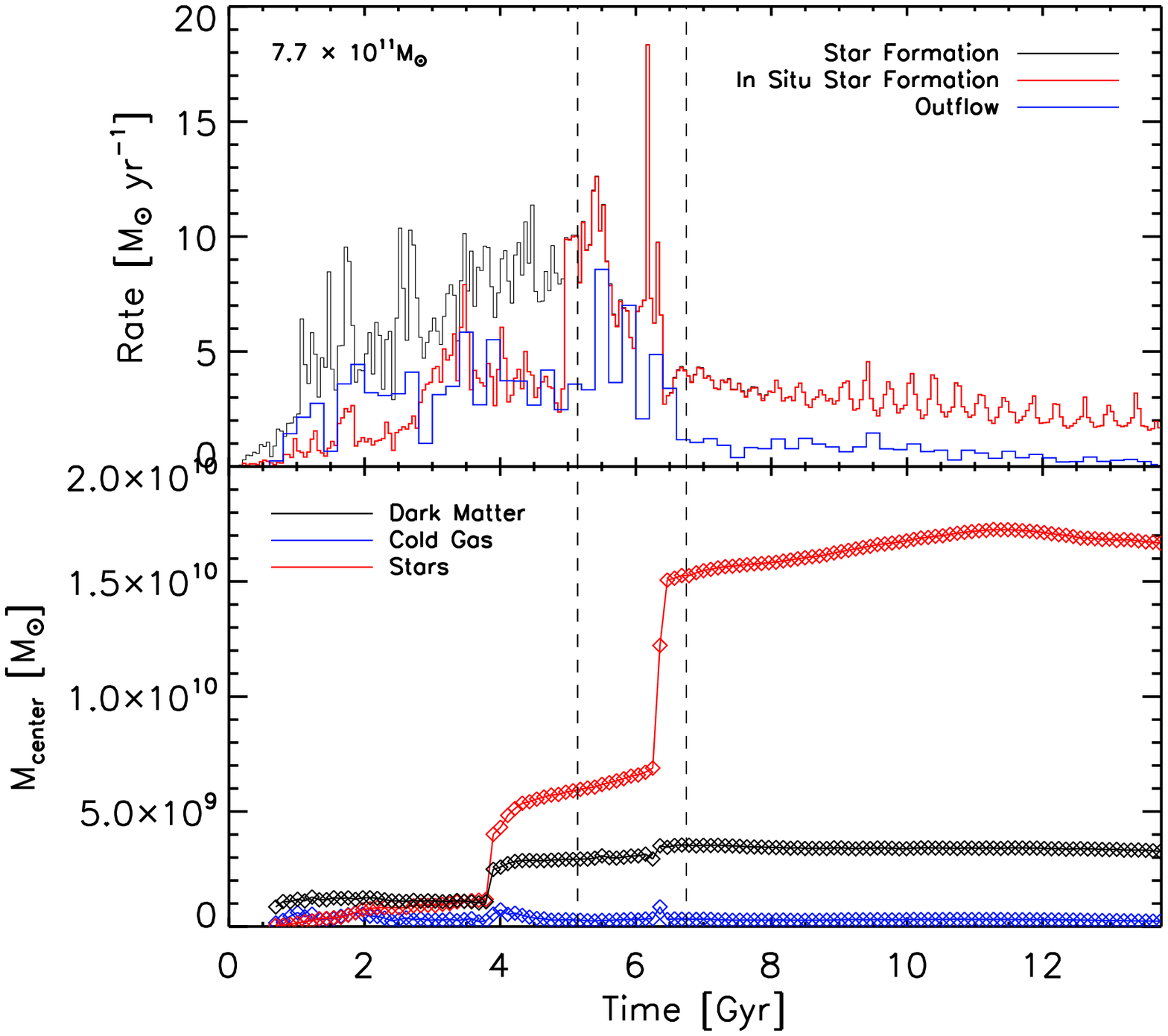}
\end{array}
$
\end{center}
\caption[Star formation and outflow history]
{
Star formation and outflow history and the evolution of the central mass concentration for four example galaxies.
In the top panels, the black curve is the total star formation history whereas the red marks the in situ star formation history for the main progenitor.
The blue curves show the history of gas outflow from the disk (gas particles are defined as outflowing if they reach a radius greater than 0.2$R_{vir}$).
The dashed vertical lines mark the beginning and end of major mergers.
The bottom panels show the evolution of the central mass concentration, i.e. mass within 0.006$R_{vir}$ at redshift zero.
Black lines indicate dark matter mass, blue the gas mass, and red the stellar mass.
For the lower mass galaxies, the merger results in a burst of star formation and, because of their high mass loading factors, an even greater mass of gas outflowing.
The result of this expulsion is a decrease in the central total mass following the merger.
In contrast, higher mass galaxies expel less mass per stellar mass formed during the merger and their central concentration dramatically increases across the merger.
}
\label{fig:mass_history}
\end{figure*}

\begin{figure}
\begin{center}
\includegraphics[width=0.5\textwidth]{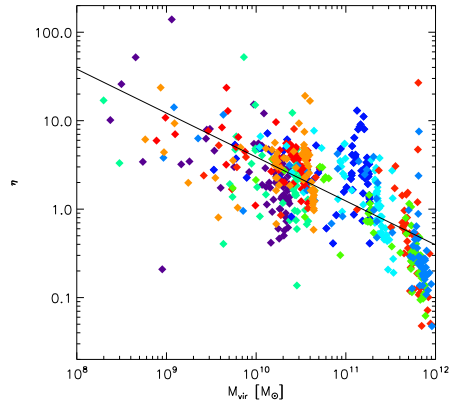}
\end{center}
\caption[Average mass loading as a function of galaxy mass]
{
Mass loading factor, $\eta$ (total gas mass ejected divided by the stellar mass formed), in 0.5 Gyr time bins  as a function of the virial mass at that time.
The colors mark different galaxies.
A power law fit to all the data points results in an exponent of -0.5, indicating that lower mass galaxies are much more efficient at driving outflows.
}
\label{fig:mass_massloading}
\end{figure}

Figure~\ref{fig:mass_history} illustrates the effect of mergers on the star formation and outflow histories of galaxies (top panels), as well as their central mass concentrations (bottom panels) for four representative galaxies with the blastwave feedback model.
Outflows were measured by using particle tracing to detect gas leaving the disk  -- outflowing material was defined to be gas particles that reached a radius greater than 0.2$R_{vir}$ after having been part of the disk.
Mergers are marked by vertical dashed lines indicating both the onset (time step where disks first show morphological distortion) and end (time of final coalescence) of the mergers.
During the mergers in all galaxies, both star formation and outflow rates peak.
However, the ratio of star formation rate to outflow rate changes with galaxy mass; in the lowest mass galaxy shown the outflow rate is four times that of the star formation rate whereas in the highest mass galaxy the outflow rate is half as much.
The mass loading factor (gas outflow rate divided by the star formation rate), $\eta$, is shown as a function of halo mass in Figure~\ref{fig:mass_massloading} for the nine galaxies simulated with blastwave feedback.
The galaxies in our sample show a -0.5 mass loading factor scaling with halo mass, which is close to that theoretically determined for the energy driven wind model.

Figure~\ref{fig:mass_history} also shows the history of the central mass of the galaxies.
The mass of stars, gas and dark matter mass within 0.006 times the redshift zero R$_{vir}$ are shown as a function of time.
This central mass can be seen as a proxy for the bulge mass.
In the two lower mass galaxies, the total central mass actually drops following the merger.
Based on the outflow rates, it is expected that the gas mass in the central region would decline.  We note, however, that the total mass in the central region declines as well.  The massive outflow originates as a hot bubble of gas that rapidly expands, and flattens the potential well \citep{Pontzen2011, Teyssier2013}.  This fluctuation in the potential well also increases the radii of the orbits of the dark matter, causing the total mass to decline, not just the gas mass.
The two more massive galaxies, however, instead undergo a dramatic increase in central material (mostly stellar mass) following the merger.
In these galaxies, the low mass loading factors results in less material being blown out of the galaxy, leading to a central stellar bulge.

\begin{figure}
\begin{center}
\includegraphics[width=0.5\textwidth]{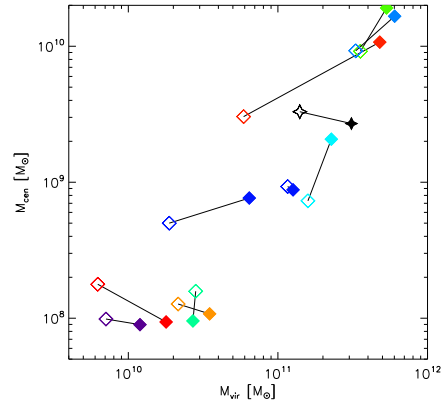}
\end{center}
\caption[Change in central mass]
{
The mass within the central region (0.006$R_{vir}$ at redshift zero) before and after a merger versus the virial mass of the galaxy at each of those times.
The open symbols show the state of the primary galaxy prior to the merger and filled symbols show galaxy post merger.
One galaxy (halo 5, seen here as the blue squares at approximately M$_{vir} = 10^{11} \Msun$) underwent two qualifying mergers and is shown twice.
The black stars represent halo 10, a galaxy from a separate cosmological simulation with the more efficient superbubble feedback model.
Galaxies with halo masses close to $10^{10} \Msun$ actually have reduced central mass following the merger whereas, except in the case of halo 10, the central mass increases for galaxies with halo masses greater than $\sim 10^{11} \Msun$.
}
\label{fig:cmass}
\end{figure}

Figure~\ref{fig:cmass} shows the change in central mass following a merger against the initial and final virial mass of the halo.
The mass trend is clear: in less massive galaxies, mergers result in lower central mass\footnote{Note that in one case, the total halo mass is slightly lower after the merger as well.  This has been seen in other works \citep{Munshi2013, Sawala2013}, too.  When the virial radius is defined at a fixed overdensity, the radius will shrink as the galaxy loses mass in the merger.  This results in a lower measured halo mass.}  while in high mass galaxies the central mass increases.
This figure also implies a transition mass with $M_{vir}\approx 10^{11} \Msun$.
Near this mass, two of the galaxies experienced central growth, whereas the other galaxy had the same central mass before and after the merger.
It should be noted, though, that the $10^{11} \Msun$ galaxy that showed almost no central mass growth also had the largest merger ratio (9.4).

The exception to these trends is halo 10, shown in black, whose central mass actually decreases following the merger, despite its large mass.
This galaxy was simulated using a the more efficient superbubble feedback model.
{\em It is remarkable for being a Milky Way-mass galaxy with one of the smallest bulge yet produced ($\sim 5\%$ of the stellar mass)} and demonstrates the potential of stellar feedback to control bulge growth during mergers.

\subsection{Discussion}
In the simulations presented here, there is a transition at $M_{vir}\approx 10^{11} \Msun$, above which mergers lead to an increase in central mass using a standard supernova feedback model.  While the mass where this transition occurs may vary with the star formation and feedback model, we discuss the broad implications of the model here.  The two simulated galaxies at $M_{vir}\approx 10^{11} \Msun$ were also examined in more detail in \citet{Christensen12a} and \citet{Christensen14}, where it was shown that the concentration of the bulges matches observed structural relations, but that there is too much early star formation compared to abundance matching results \citep[e.g.,][]{Moster2012}. The bulges overall are too massive (by a factor of $\sim$4 or more assuming a mass-to-light ratio of 1 in the $H$-band).  
We also include an additional Milky Way-mass galaxy simulated with the more efficient superbubble feedback model.
This galaxy not only has a reduced central concentration following a major merger, it also has a very low B/T of $0.05$.
This galaxy demonstrates the potential for stellar feedback to limit bulge growth but it will require much more extensive analysis before firm conclusions can be drawn.

There are two potential solutions to reducing the bulge mass in simulated galaxies with $M_{vir} \gtrsim 10^{11} \Msun$. 
First, if high-redshift star formation were decreased in simulations, then gas fractions would be larger in high-$z$ disks, causing mergers to be less efficient at building bulges.
Observations show that galaxies at $z > 2$ have higher gas fractions than in similar mass galaxies at $z = 0$ \citep{Tacconi2010}.
The gas-rich galaxies observed at high $z$ are progenitors of more massive galaxies than we study here.  
Nonetheless, at $z = 0$ gas fractions increase as galaxy mass decreases \citep[e.g.,][]{Geha06, Dutton2011}.  Assuming this trend also holds at high $z$, the progenitors of Milky Way-mass galaxies would be more gas-rich than today, making high-redshift mergers relatively less efficient at bulge formation.  Given that simulations are already known to produce excessive high-redshift star formation rates, it is likely that they have tended to be too gas-poor at high $z$, leading to larger bulges than are realistic.  Those simulations that match the high $z$ evolution of the stellar mass-to-halo mass relation \citep[e.g.,][]{Brook2012d, Aumer2013, Hopkins2013d} may do better at making smaller bulges.  Indeed, \citet{Aumer2013} showed that their bulges were smaller, but \citet{Aumer2014} showed that the growth of these central regions was still inconsistent with observations, suggesting that matching the stellar mass-to-halo mass relation is not by itself a sufficient criterion to ensure realistic bulge growth.  Other simulations have yet to examine bulge growth.

As discussed in Section 4.3, inclusion of feedback from young stars (before they produce supernova) has been shown to bring high redshift simulated star formation rates in line with abundance matching results \citep{Stinson13, Aumer2013}.  
However, the inclusion of this feedback at the levels currently required to reproduce observed properties leads to too thick and extended of stellar disks \citep{Roskar2013, Aumer2014}.  The simulations we have examined in this section lack AGN feedback, as have most of the simulations that examine central mass growth.  AGN feedback could be a natural contender to provide extra feedback.  Additional motivation for AGN feedback can be found in \citet{Dutton2011}, who showed that dark matter halos that host late-type Milky Way-mass galaxies seem to require some halo expansion (assuming a non-evolving Chabrier IMF) in order to match the zero point of the Tully-Fisher relation.  Halo expansion at these masses is not reproduced in current simulations that include early stellar feedback \citep{DiCintio2013}, but AGN feedback could create additional fluctuations in the gravitational potential wells of these massive galaxies that would expand the dark matter orbits \citep{Martizzi2013} 

AGN feedback could be beneficial in two ways; it could regulate star formation across the entire galaxy, potentially bringing simulated star formation rates into agreement with observations, and it could lead to more blowout of low angular momentum gas that can further shrink the sizes of bulges in massive disk galaxies. 
For instance, it has been been pointed out that there is an intriguing match between the global star formation history of the Universe and the accretion history of black holes \citep{Terlevich1998,Silverman2008}, indicating that AGN feedback may indeed to tied to star formation regulation in galaxies.  
Unfortunately, the modeling of AGN feedback is highly uncertain, in particular how it couples with the surrounding gas.
AGN feedback models can be broadly classified between those dominated by energy injection \citep[e.g.][]{Springel2005} and those dominated by momentum injection \citep[e.g.][]{DeBuhr2010}.
In the latter case, the velocity of winds has a profound effect on the range over which AGN limit star formation.
Low velocity winds (as is typical when the momentum is generated by radiation pressure) affect only the centers of galaxies \citep{DeBuhr2010} while high velocity winds (like those arising from accretion disks) can affect star formation over a much larger extent \citep{DeBuhr2012} and, potentially, the growth of bulges in Milky Way-mass galaxies (Choi, private communication).

One strong appeal of AGN feedback is that it might remove gas from the galaxy at early times that could later be re-accreted and contribute to star formation at low $z$, potentially resulting in more realistic star formation histories.
While many simulations form too many stars at high $z$, an additional consequence is that they underproduce stars at $z < 1$ \citep{Somerville2014}.  
Moreover, unlike supernova feedback, AGN feedback can be somewhat independent from the mass of stars formed, which gives it a greater amount of latitude to affect scaling relationships.
Given that feedback from young stars combined with supernovae has yet to fully satisfy observational constraints (e.g., simultaneously match stellar to halo mass relations, disk thickness, and bulge/disk size growth), AGN feedback is looking more and more appealing as a potential regulator of star formation.

Perhaps the most attractive aspect of AGN feedback, though, is that it provides a mechanism that may preferentially remove gas from the centers of galaxies.  
Despite the recent successes of cosmological simulations in making extended disks, the B/T ratios always seems to reside on the high end of what is observed \citep[e.g.,][]{Christensen14}.  
As described at the beginning of this section, mergers in dwarf galaxies lead to a burst of star formation that drive outflows of low angular momentum gas.  
Are AGN the missing mechanism to reduce bulge mass in higher mass galaxies by removal of low angular momentum gas?  
In an ideal scenario, mergers at high redshift might drive gas to galaxy centers, leading to fueling of a black hole and the resultant feedback would reduce bulge growth by expelling excess gas from the center of the galaxy.
As an additional bonus, AGN feedback may be able to drive gas at higher velocities than supernova feedback, allowing AGN to be more efficient at removing gas despite deeper potential wells in more massive galaxies.

However, AGN feedback cannot be a panacea.
While potentially critical to limiting classical bulge growth, AGN feedback does not explain the presence of massive galaxies that completely lack a classical bulge.
Based on the M$_{\textrm{BH}}$-sigma relation, galaxies with small classical bulges also have small black holes, which would be less effective at removing low angular momentum gas through feedback.
At the far extreme, galaxies with only a pseudobulge do not lie on the M$_{\textrm{BH}}$-sigma relation \citep{Kormendy2013} and tend to have small bulges, if at all.
AGN feedback, therefore, has the potential to scale down the mass of classical bulges but to explain the existence of massive, pure-disk galaxies, stellar feedback is likely critical.

\section{Summary and Future Prospects} 
\label{sec:5}
 
Simulators face a number of hurdles in studying the formation of bulges in a cosmological context.
Simulations must include large scale structure to capture tidal torques and properly model the angular momentum build-up of galaxies, but to resolve bulges they must also have very high force resolutions.  Covering this range of scales is exceptionally computationally expensive. 
Convergence of the inner few 100 pc will still remain a challenge for the foreseeable future, however, state-of-the-art simulations with sub-100pc force resolution are now allowing studies of these inner regions that were previously impossible.  Until recently, such resolution could only be adopted in idealized merger simulations.  Hence, simulators are now in a position to study bulge formation in a fully cosmological setting for the first time.

Much progress has been made: \citet{Guedes2013}, \citet{Okamoto2013} and \citet{Aumer2013} were able to form bulges with low bulge-to-total ratios while \citet{Christensen14} produced bulges that matched observed scaling relations.
However, the B/T ratios of most cosmological simulations still tend to cluster on the high side of what is observed in disks at comparable stellar masses and the bulges are frequently overly-concentrated.
Moreover, no simulation has yet showed realistic bulge formation in conjunction with a realistic star formation history.  It is usually assumed that more feedback, particularly at high redshift, can lower bulge masses and reduce their concentration. First, suppressing star formation overall at high redshift will lead to higher gas fractions, which lowers the efficiency of bulge formation in mergers \citep{Hopkins2009a}.  Second, feedback can prevent early star formation while increasing later star formation rates through the reaccretion of previously-ejected gas, in better agreement with derived stellar-to-halo mass relations.  Third, feedback can eject low angular momentum gas entirely, reducing the overall size of bulges and bringing them into line with observations \citep[e.g.,][]{VandenBosch2001c}.

However, the form of the feedback that solves this problem remains elusive.  Supernova feedback alone leads to too much early star formation \citep{Stinson13} and increasing it without limit results in blown-apart galaxies \citep{Agertz2013}.  Including feedback from young stars (UV ionization, radiation pressure, momentum injection in winds) enables the reproduction of observed stellar-to-halo mass trends \citep{Agertz2013, Hopkins2013d, Aumer2013}, however these galaxies have too thick of disks \citep{Roskar2013} and their disk growth is weighted too much to the outer disks \citep{Aumer2014}. 
As discussed in this chapter, a new model for superbubble feedback shows great promise in its ability to reduce bulge growth but the resulting galaxies require much more extensive evaluation before it is shown to be viable.
We are left to wonder if the adopted sub-grid feedback models are the problem, or whether stellar feedback alone simply cannot satisfy all observational constraints.  If the latter case, AGN feedback may be invoked to contribute as well.

In addition to the problem of massive bulges, simulations must also contend with the apparent over-production of classical bulges in CDM.
Can feedback reduce the formation of classical bulges in significant enough numbers to reproduce observational trends?  
It has been shown that classical bulges are in the minority in the Local Volume \citep{Weinzirl2009,Kormendy2010} and some disk galaxies as massive as the Milky Way appear to be either bulgeless or dominated by pseudobulges.  Yet simulations have clearly shown that mergers can create classical spheroids, either through a redistribution of stars in gas-poor mergers, or by inducing gas flows to the center of galaxies that subsequently form stars.  Given the ubiquity of mergers in CDM, the lack of classical bulges in galaxies is a serious problem for CDM galaxy formation theory \citep{Peebles2010}.  If we wish to affirm the CDM model, we are left with two choices: either some mergers must form pseudobulges instead of classical bulges, or mergers must tend to {\it oppose} bulge formation so that some other mechanism such as disk instabilities forms the majority of bulges instead.

Both of these scenarios may be at play to some degree.  Minor mergers have been seen to induce bars that create bulges with small $n$ in cosmological simulations \citep{Guedes2013, Okamoto2013}. Idealized major mergers have also been shown to lead to small $n$ bulges \citep{Keselman2012}.  These results suggest that a picture in which all mergers lead to classical bulge formation is not complete. 

Yet even if some mergers led to pseudobulge formation, current theoretical models would still predict more massive bulges than are observed.  This disagreement suggests that some mechanism is suppressing bulge formation.  The formation of bulgeless disk galaxies, wherein mergers can fuel the {\it loss} of bulge material, offers an intriguing case study that may be extended to more massive galaxies.  For instance, it is likely that feedback was more effective at high $z$, when the potential wells of galaxies are shallower \citep{Brook2012c}.  Furthermore, some simulations suggest that mass loading factors could have been greater at high redshift \citep{Muratov2015}. Could the progenitors of modern-day L$^{\star}$ galaxies have had high enough mass loading factors to prevent bulge formation during high-redshift mergers?
Either AGN feedback or new models of stellar feedback may be able to fuel strong enough galactic winds to remove enough low angular momentum material.

Within this picture, we can imagine two phases of bulge evolution for a galaxy like our own Milky Way with a relatively quiescent merger history \citep{Hammer2007}. At high-$z$ during the era of major mergers, the gas fraction of the galaxy is high.  This reduces the size of the bulge that can be formed in mergers, and merger-induced outflows are also more efficient because the potential well is shallower.   The combination of these two facts could suppress classical bulge formation at high $z$.  At lower $z$, secular processes or bars induced by minor mergers could create bulges that look like pseudobulges \citep{Shen2010}.  

M31 represents the opposite extreme with its large classical bulge \citep{Kormendy2010}. 
If high gas fractions and outflows are acting to reduce bulge formation at early times, this suggests a late major merger when gas fractions were lower induced the formation of its classical bulge.  
Such a lower $z$ massive merger is in agreement with trends observed in M31's stellar halo \citep{Deason2012b, Gilbert2014}.  

The contrast between a given massive spiral galaxy that contains a classical bulge, such as M31, and one that does not, such as the Milky Way or M101, may be explained by stochasticity in merger histories.
However, the observed correlation between classical bulges and high-density environments requires a more general explanation.
For instance, as discussed in greater detail in Kormendy's summary chapter, in dense environments galaxies are more likely to undergo mergers and gas that could otherwise be accreted to regrow a large disk remains suspended as hot, X-ray emitting gas.
In low density environments, the galaxies experience fewer mergers, what mergers they do experience are more gas rich, and they are able to continually accrete additional gas from the cosmic web.
In this scenario, stellar and AGN feedback could limit bulge-growth through mergers in general and especially at low masses while differences in the environment could account for the three orders of magnitude overlap in mass of disk and elliptical galaxies.

Unfortunately, the computational expense of simulating a broad sample of galaxies at resolutions high enough to resolve the bulge has made it extremely difficult to computationally study the connection between morphology and environment.
Cosmological simulations that focus on individual galaxies are just beginning to be able to resolve bulges but there are far too few galaxies for population studies.
In these simulations, the initial conditions are instead chosen to produce galaxies of the desired morphology.  Generally, a quiescent merger history is picked to increase the probability of a galaxy with a low B/T forming.  Additionally, embedding high resolution simulations in a dense environment increases the number of high resolution particles required, and drives up the computing costs.

Cosmological simulations of a volume of space (as opposed to a specific galaxy) can achieve larger numbers of galaxies but at the cost of resolution.
For instance, \citet{Snyder2015} was able to roughly reproduce the relationship between galaxy morphology and overdensity but with a gravitational softening length of 710 pc, which is larger than most bulges and certainly insufficient for distinguishing pseudobulges from classical bulges.
Tantalizingly, \citet{Sales2012} identified the alignment between the angular momentum of the accreting baryons and the galaxy disk as being a key factor in the morphology of galaxies.
In galaxies where there was misalignment between the accreting gas and the disk, the net rotation of the galaxy was reduced and it was more likely to be a spheroid.
In contrast, if the angular momentum of the accreting gas and existing disk had similar alignment,  the galaxy was more likely to be a disk. Once again, though, the resolution was comparatively low (softening lengths of 500 pc) and these simulations are known to produce too many stars \citep{Crain2009} and, seemingly, too many spheroids.

SAMs offer yet another avenue to examine the connection between galaxy morphology and environment and many of them have been able to reproduce the approximate environmental dependency \citep[e.g.][]{Baugh1996,Somerville1999,Cole00}. However, recall that SAMs generally do not distinguish between classical and pseudobulges, limiting the reach of these results.
Additionally, SAMs have typically been based off of idealized binary merger simulations, which tend to have much lower amounts of stellar feedback than cosmological simulations have found necessary.
As SAMs become more nuanced in their modeling of bulges and as the results from larger samples of high-resolution cosmological simulations are analyzed and implemented into them, they will become all the more important for connecting galaxy morphology to environment.

Is it possible for bulge formation to be compatible with merger rates?  Can we identify a redshift range over which the fraction of galaxies with classical bulges can be reproduced while simultaneously matching the small fraction of mass in pseudobulges?  This remains to be seen.  However, it is clear that mergers cannot simply form bulges as historically believed if CDM is the correct model.  Despite the fact that a proper treatment of gas fractions can reproduce the number density of observed spheroids for $L^{\star}$ galaxies and smaller, these bulges always tend to reside on the massive side compared to observations and include too large a fraction of classical bulges.  We must explore other options, and feedback-driven outflows generated during mergers are a natural choice for reducing the bulge formation.

\comment{
\begin{acknowledgement}
\comment{
If you want to include acknowledgments of assistance and the like at the end of an individual chapter please use the \verb|acknowledgement| environment -- it will automatically render Springer's preferred layout.
}
\end{acknowledgement}
}

\end{document}